# Topological, non-topological and instanton droplets driven by spin-transfer torque in materials with perpendicular magnetic anisotropy and Dzyaloshinskii–Moriya Interaction

Mario Carpentieri[1*], Riccardo Tomasello[2], Roberto Zivieri[1,3], Giovanni Finocchio[4]

[1]Department of Electrical and Information Engineering, Politecnico di Bari, via E. Orabona 4, I-70125 Bari, Italy

[2]Department of Computer Science, Modelling, Electronics and System Science, University of Calabria, via P. Bucci 41C, I-87036, Rende (CS), Italy

[3]Department of Physics and Earth Sciences and CNISM Unit of Ferrara, University of Ferrara, Ferrara, via Saragat 1, I-44122, Ferrara, Italy

[4]Department of Electronic Engineering, Industrial Chemistry and Engineering, University of Messina, c.da di Dio, I-98166, Messina, Italy

*Correspondence to mario.carpentieri@poliba.it

The interfacial Dzyaloshinskii–Moriya Interaction can modify the topology of droplets excited by a localized spin-polarized current. Here, we show that, in addition to the stationary droplet excitations with skyrmion number either one (topological) or zero (non-topological), there exists, for a fixed current, an excited mode with a non-stationary time behavior. We call this mode "instanton droplet", which is characterized by time domain transitions of the skyrmion number. These transitions are coupled to an emission of incoherent spin-waves that can be observed in the frequency domain as a source of noise. Our results are interesting from a fundamental point of view to study spin-wave emissions due to a topological transition in current-driven systems, and could open the route for experiments based on magnetoresistance effect for the design of a further generation of nanoscale microwave oscillators.

Solitons are self-localized wave packets that can be observed in media characterized by non-linear and dispersive constitutive laws and can be also classified on the basis of their topology, i.e. the skyrmion number[1,2,3,4,5]. The study of magnetic solitons[6,7] has been reinforced by the experimental evidence that spin-transfer torque (STT)[8] can nucleate both static and dynamic solitons from a uniform state or manipulate strongly non-uniform magnetic configurations. In particular, static solitons can be achieved in either in-plane (vortices, antivortices, and domain walls) or out-of-plane devices (bubbles, skyrmions, and domain walls) and they can behave like particles, e.g. vortex[9] and skyrmion[10,11,12,13] can rotate around a nanocontact in point contact geometries (oscillators), skyrmions and domain walls can be shifted in a track (racetrack memory)[14,15,16,17]. On the other hand, dynamical solitons, that are unstable in dissipative magnetic materials, can be sustained by the STT and used as source of microwave emissions (self-oscillations) in nanoscale oscillators[18,19,20,21,22,23]. In the recent literature, much attention has been given to the investigation of dynamical properties of topologically protected particles in materials with perpendicular magnetic anisotropy, i.e. skyrmions[24], while only few works have been devoted to the identification and study of topological modes[25]. In detail, the excitation of a topological mode was predicted by Zhou *et al.*[25] and used to explain recent experimental observations in CoNi[26].

In this paper, we describe how the transition from non-topological droplet (NTD) to topological droplet (TD) or dynamical skyrmion occurs as a function of the *i*-DMI[27,28,29]. The adjectives "non-topological" and "topological" are used to identify a mode with a skyrmion number zero ($S = 0$) and one ($S = 1$), respectively, since the skyrmion number characterizes the topological structure[20]. Our key result is that the mode transition from NTD to TD is achieved through an intermediate *i*-DMI region, where the time-domain traces of the excitation exhibit non-stationary changes from NTD to TD and vice versa. We have called this new mode "instanton droplet" (ID), which can be seen as a combination of the TD and NTD, characterized by continuous changes of the skyrmion number. This excitation can be experimentally observed by means of microwave emissions data.

**Results**

**Phase diagram description.** Fig. 1 shows a sketch of the studied device, consisting of a spin-valve of Pt(5 nm)/Co(0.6 nm)/Cu (4 nm)/CoPt(4 nm). The ultrathin Co layer acts as free layer (FL) (square cross section of 400 × 400 nm$^2$ and thickness of 0.6 nm), while the CoPt acts as fixed layer or polarizer and it is shaped as a circular point contact (diameter $d_c$ = 70 nm) in order to locally inject the current into the FL. Both FL and polarizer have an out-of-plane magnetic state at zero bias field. A Cartesian coordinate system, with the *x*- and *y*-axes oriented along the in-plane directions of the device and the *z*-axis along its thickness, has been introduced.

Fig. 2 summarizes the phase diagram of the magnetization as a function of the current density (swept back and forth) and the *i*-DMI parameter *D*. For the sake of simplicity, in the following, the current density *J* < 0 (the current flows from the FL to the fixed layer) is given in modulus. Five different states can be identified, two static states: uniform state along the *z*-direction (FM) and static skyrmion (SS), and three dynamical states, NTD[20, 21], TD and ID. The *i*-DMI regions described above are separated by straight horizontal lines, as obtained with a *D* resolution of 0.05 mJ/m$^2$, whereas the dotted line for *D* = 3.7 mJ/m$^2$ marks the *i*-DMI value above which the skyrmion, once nucleated, is stable without current. In some current regions, the FM state is overlapped with the three dynamical states, since all the modes deal with a sub-critical Hopf bifurcation[30] (finite power at the threshold, current density hysteresis, the mode is switched off at a current density smaller than the excitation value, and the oscillation axis is different from the equilibrium configuration in the FM state). Starting from the FM state, the modes are excited at *J* = 7.0 × 10$^7$ A/cm$^2$ (*D* < 3.0 mJ/m$^2$) independently of the *i*-DMI, while the switch-off current density depends on the *i*-DMI. This is due to the dependence of the *i*-DMI field (see equation (4) in Methods) on the spatial derivative of the magnetization that does not influence the excitation current in the uniform state. The SS state is achieved for 1.6 < *D* ≤ 3.0 mJ/m$^2$ from the TD state with a reversible transition, while a static skyrmion, once excited, is stable with no current for *D* ≥ 3.7 mJ/m$^2$. From the theory developed in [29], the critical *i*-DMI parameter which stabilizes the skyrmion state is given by $D_c = 4\frac{\sqrt{Ak_{eff}}}{\pi} = 4.4$ mJ/m², where $k_{eff} = k_u - \frac{1}{2}\mu_0 M_s^2$. As expected and already well discussed in[29], the analytical value is larger than the micromagnetic one because the $D_c$ expression has been derived within a 1D model.

We have also performed a systematic study of the stability phase diagram as a function of different physical parameters, namely the perpendicular anisotropy constant $k_u$, the exchange stiffness constant $A$ and the Gilbert damping $\alpha_G$. We found that, in order to achieve magnetization dynamics at zero external magnetic field, a high perpendicular anisotropy ($k_u > 0.95$ MJ/m³) is needed, whose values have been already measured in similar devices[31].

## Discussion

**Topological and non-topological droplet.** The NTD mode is excited for $D \leq 0.5$ mJ/m² (see supplementary material MOVIE 1 for the NTD mode at $D = 0$ and $J = 8.5 \times 10^7$ A/cm²). The NTD dynamics concerns a 360° in-plane rotation of domain wall spins[18, 20, 21]. Unlike previous studies, where the NTD is characterized by two or four regions[19, 20] of opposite topological density, here the NTD exhibits a more complex behavior as can be also seen from a snapshot of the topological density in Fig. 4a and from the supplementary material MOVIE 2 ($D = 0$ and $J = 8.5 \times 10^7$ A/cm²). Close to $D = 0.25$ mJ/m², the NTD is characterized by a small shift of the droplet core together with the domain wall spins rotations (see supplementary material MOVIE 3 for the magnetization dynamics at $D = 0.25$ mJ/m² and $J = 8.5 \times 10^7$ A/cm²). The origin of the different topological density of the NTD will be discussed ahead in the last subsection.

The TD is excited for $1.6 < D \leq 3.0$ mJ/m². It exhibits a core breathing dynamics that is synchronized with a 360° in phase rotation (space and time) of the domain wall spins (see supplementary material MOVIE 4 for $D = 2.5$ mJ/m² and $J = 8.5 \times 10^7$ A/cm²), which can be seen as a continual change from Néel (radial outward and inward) to Bloch (counter clockwise and clockwise) skyrmion magnetic texture ($S = -1$, see a snapshot of the topological density in Fig. 4a and the supplementary material MOVIE 5 for $D = 2.5$ mJ/m² and $J = 8.5 \times 10^7$ A/cm²).

Fig. 3a and b show the frequency-current and the output power vs current for $D = 2.0$ mJ/m² as computed from simulations, respectively. The oscillation frequency of the TD (at values smaller than the ferromagnetic resonance (FMR) frequency, about 37 GHz) decreases with increasing current[19]. The oscillation power computed from the z-component of the magnetization (perpendicular polarizer) shows a finite value at the excitation current when the initial state is the FM, as expected for a sub-critical Hopf bifurcation, whereas it tends to zero as soon as the SS state is approached. These results show that the TD state can be seen as a reversible linear mode of the SS justifying the name of dynamical skyrmion already used in literature.

Looking at the spatial distribution of the oscillating spins of the TD modes (see Fig. 3c) for the two-dimensional profile of the TD for $D=2.5$ mJ/m$^2$ and $J = 8.5 \times 10^7$ A/cm$^2$ as computed with the micromagnetic spectral mapping technique[32, 33], it is easy to demonstrate that the dimensionless output power $p$, due to a variation in the Giant Magnetoresistance (GMR) signal, is related to the breathing mode as:

$$p = \begin{cases} \left(1 - \dfrac{r_{min}^2}{r_{max}^2}\right) & r_{max} \leq r_c \\ \left(1 - \dfrac{r_{min}^2}{r_c^2}\right) & r_{max} > r_c \end{cases} \quad (1)$$

being $r_{min}$ and $r_{max}$ the minimum and maximum radius of the TD during the breathing and $r_c$ the radius of the point contact (see Fig. 3c). The dimensionless output power calculated according to equation (1) shows a very good agreement with the one computed by means of micromagnetic simulations (Fig. 3b).

The TD mode described in this study, already observed in[25], is different from the ones seen in[17] where the breathing mode is just a transient due to the application of the spin current, in[34] where the breathing mode is a resonant state of a static skyrmion excited by a microwave field, and in[26] where no synchronization between the core breathing and domain wall spins precessions is detected. Other topological modes have been already found in in-plane materials such as vortex-antivortex pairs[35] ($S = \pm 1$) and vortex-quadrupole ($S = \pm 2$)[36].

**Instanton droplet.** In general, in 2D systems the topological states are represented by local minima in the free energy landscape separated by a finite energy barrier. This energy barrier is proportional to the exchange contribution in a texture where the spins are not locally parallel but resemble through a hedgehog-like spin configuration with a singular point in the middle. This magnetic structure can be called "instanton", which in our framework is a time-dependent magnetization configuration connecting different topological states[37, 38]. In the *i*-DMI region $0.5 < D \leq 1.6$ mJ/m$^2$, time domain changes in the topology (skyrmion number) of the topological mode are found, in particular a configuration linkable to NTD or to TD can be observed.

The first definition of instanton was classical and referred to localized finite-action solutions of the classical Euclidean field equations with finite Euclidean action[38]. Recently, an instanton dynamics has been introduced to study the quantum dynamics of vortices in magnetic disks starting from the generalized Thiele's equation[39]. In analogy, we have defined an instanton as a time-dependent configuration connecting different topological states. In other words, in our framework the term "instanton" has been introduced with the aim to extend this important notion to low-dimensional semi-classically described magnetic systems, thus achieving the classical correspondence of the pseudoparticles theoretically found in[40]. This has been obtained by solving micromagnetically the LLGS equation of motion for low-dimensional magnetic systems. Specifically, we identify with "instanton droplet" only the $i$-DMI region where the dynamics is characterized by a variation in time of the topological charge (skyrmion number) passing from TD ($S = -1$) to NTD ($S = 0$) and vice versa.

The continual changes in the droplet topology generate incoherent emission of spin waves[41,42], being the topological transition non periodic. The magnetization dynamics for $J = 8.5 \times 10^7$ A/cm$^2$ at three different values of $D$ (0.75, 1.00, and 1.25 mJ/m$^2$) can be seen in the supplementary material MOVIES 6-8. To highlight the main characteristics of the dynamical states of the phase diagram, the Fourier spectra for different values of $i$-DMI ($D$ = 0.25, 0.75, 1.00, and 2.50 mJ/m$^2$) are shown in Fig. 4b. The NTD and TD are characterized by a single mode (the Fourier spectra exhibit a main frequency peak at 4.12 GHz for $D$ = 0.25 mJ/m$^2$ and at 3.17 GHz for $D$ = 2.5 mJ/m$^2$, respectively). In the ID region, the time-domain non-stationary topological transitions give rise to the excitations of incoherent spin waves leading to noisy Fourier spectra (see the spectra for $D$=0.75 and $D$=1.00 mJ/m$^2$). From an experimental point of view, it is possible to detect the ID region by performing microwave emission measurements.

Fig. 5 summarizes the oscillation frequency linked to the mode with larger power as a function of the $i$-DMI for a fixed current density of $8.5 \times 10^7$ A/cm$^2$, as indicated in Fig. 2. The transition from a dynamical region to the other one is similar to the topological transitions from skyrmion to a uniform ferromagnetic state or vice versa as described in[43]. In particular, the topological changes characterizing the instanton droplet can be seen as finite-time singularities[43] which are driven by the spin-polarized current. The inset of Fig. 5 shows the energy related to the continual change of the topology as a function of $D$, within a time window of 42 ns. The energy increases with $D$ because more stable topological droplets are achieved.

**Control of the topological density of NTDs.** Finally, we have studied the mode excitation for different contact diameters. Fig. 6 shows the Fourier spectra of the magnetization dynamics when $J = 8.5 \times 10^7$ A/cm$^2$ and zero DMI for three different values of $d_c$. As expected, by injecting the same current density, the frequency of the domain wall rotation increases while decreasing the nanocontact diameter due to energy balance between the current input and the Gilbert damping. Moreover, for the smallest $d_c$, the spatial distribution of the topological density is composed of four different regions (see supplementary material MOVIE 9 for $d_c$ = 40 nm), whereas for the larger diameters a more complicated topological density distribution is obtained, including several alternated regions with positive and negative topological density (see supplementary material MOVIE 10 for $d_c$ = 100 nm). Our results show that the topological density of the NTD can be controlled by the size of the nanocontact $d_c$.

Micromagnetic simulations show qualitatively the same phase diagrams for contact size larger than 50 nm. In particular, for $D$ = 2.50 mJ/m$^2$, the TD state is excited in the range (3.0 ≤ $J$ ≤ 6.5) × 10$^7$ A/cm$^2$ and (7.0 < $J$ ≤ 9.0) × 10$^7$ A/cm$^2$ for $d_c$ = 100 nm and $d_c$ = 70 nm, respectively. When $d_c$ = 40 nm, the TD mode is not excited, even when the spin-polarized current is increased. This is due to the fact that the small size of the nanocontact hampers the breathing mode of the droplet. However, at reduced contact diameter sizes $d_c$ < 50 nm, to stabilize a TD, an external out of plane field is necessary (not shown).

In summary, micromagnetic results point out that the additional degree of freedom of the i-DMI energy together with a spin-polarized current can drive transitions from either static to dynamical or dynamical to dynamical states, also implying a change in the topology during these transitions[43]. These results indicate a route for the fundamental study of topological transitions in driven systems. Moreover, we have identified a new mode we called "instanton droplet", which is characterized by time domain transitions of the topological charge.

Finally, the breathing mode of the TD can be used as basis for the design of high power spin-transfer torque oscillators by considering a redesign of this system in a magnetic tunnel junction[44], where it is possible, in a phenomenological way, to link the microwave output power and the size of the breathing mode.

## Methods

**Micromagnetic model**

The numerical results presented in this paper are based on micromagnetic simulations as computed from the Landau-Lifshitz-Gilbert-Slonczewski (LLGS) equation[8]

$$\frac{d\mathbf{m}}{d\tau} = -\mathbf{m} \times \mathbf{h}_{\text{eff}} + \alpha_G \left( \mathbf{m} \times \frac{d\mathbf{m}}{d\tau} \right) + B\,\mathbf{m} \times (\mathbf{m} \times \mathbf{p}) \qquad (2)$$

where $\mathbf{m}$ and $\mathbf{h}_{\text{eff}}$ are the normalized magnetization and the effective field of the ferromagnet. $\tau = \gamma_0 M_s t$ is the dimensionless time. $\gamma_0$ is the gyromagnetic ratio and $\alpha_G$ is the Gilbert damping. The third term in equation (2) is the dimensionless Slonczewski spin-transfer torque, where $\mathbf{p} = \mathbf{P}/M_s$ represents the dimensionless magnetization in the fixed layer, and $B = \frac{g\mu_B}{\gamma_0} \frac{J}{M_s^2 ed} P(\mathbf{m},\mathbf{p})$, where $g$ is the Landè factor, $e$ and $\mu_B$ are the electric charge and the Bohr magneton, respectively. $J$ is the current density and $d$ is the thickness of the FL, while $P(\mathbf{m},\mathbf{p})$ is a polarization function depending on the relative orientation of the magnetizations. The effective field includes the standard magnetic field contributions and the DMI field[48]. A complete description of the numerical framework can be found in[45, 46, 47].

The coupled Pt heavy metal adds the *i*-DMI as an additional degree of freedom in the energy landscape of the FL. The *i*-DMI energy density and the effective field expressions are[48]:

$$\varepsilon_{i-\text{DMI}} = D\left[m_z \nabla \cdot \mathbf{m} - (\mathbf{m} \cdot \nabla)m_z\right] \tag{3}$$

$$\mathbf{h}_{i-\text{DMI}} = -\frac{1}{\mu_0 M_S}\frac{\delta \varepsilon_{i-\text{DMI}}}{\delta \mathbf{m}} = -\frac{2D}{\mu_0 M_S}\left[(\nabla \cdot \mathbf{m})\hat{z} - \nabla m_z\right] \tag{4}$$

respectively, where $m_z$ is the z-component of the normalized magnetization, $M_s$ is the saturation magnetization of the FL, $\delta$ stands for functional (variational) derivative and the ultrathin film hypothesis $\left(\frac{\partial \mathbf{m}}{\partial z} = 0\right)$ is considered, while the boundary condition is $\frac{d\mathbf{m}}{dn} = \frac{1}{\xi}(\hat{z} \times \mathbf{n}) \times \mathbf{m}$, where $\xi = \frac{2A}{D}$ (being *A* the exchange constant) is a characteristic length in the presence of DMI[29].

The micromagnetic parameters, typical of Co, are saturation magnetization $M_s$ = 900 kA/m, perpendicular uniaxial anisotropy constant $k_u$ = 1.10 MJ/m³, exchange stiffness constant $A$ = 20 pJ/m, and magnetic damping $\alpha_G$ = 0.1. All the simulations have been performed without any bias field and for a range of *i*-DMI parameter $0 < D < 4.0$ mJ/m² and of current density $0 < J < 13 \times 10^7$ A/cm².

The topological density *n* has been computed directly from the magnetization configuration as $n = \mathbf{m} \cdot (\partial_y \mathbf{m} \times \partial_x \mathbf{m})$, while the skyrmion number *S* is given by $S = \frac{1}{4\pi}\int n(x,y)\,dx\,dy$. Here, the droplet is considered topological or non-topological if $|S| \approx 1$ or $S \approx 0$, respectively, for all *t*, while the instanton droplet is characterized by a time dependent skyrmion number $|S(t)|$, changing from 0 to 1.

# Figure captions

**Figure 1 | Device under investigation.** Spin valve with a point contact geometry, where the Co free layer is coupled to the Pt underlayer. For the sake of clarity, an enlarged view of the nanocontact (fixed layer) and the diameter $d_c$ of the nanocontact are illustrated, together with the dimensions $l \times l$ of the square cross section.

**Figure 2 | Magnetization phase diagram.** Stability phase diagram of the magnetization ground-state as a function of the modulus of the current density and of $D$ at zero external magnetic field. Letters A, B, C and D are linked to Fig. 4b. The meaning of the symbols in the phase diagram are as follows. FM: ferromagnetic; SS: static skyrmion; TD: topological droplet; NTD: non-topological droplet, ID: instanton droplet.

**Figure 3 | Oscillation frequency and power vs. current density. a**, Frequency-current density hysteresis loop at $D = 2.0$ mJ/m$^2$: the black (red) arrows indicate the path where the initial state is FM (SS). **b**, Output power as a function of current density for $D = 2.0$ mJ/m$^2$. The solid line refers to the analytical computation from equation (1), while the dotted line is determined by micromagnetic calculations. The black (red) arrows indicate the path where the initial state is FM (SS). **c**, Two dimensional spatial profile of the TD for $D=2.5$ mJ/m$^2$.

**Figure 4 | Topological density and Fourier spectra. a**, Spatial distribution of the topological density (a color scale is represented, red +1, blue -1) for the three dynamical states NTD, TD, and ID at $i$-DMI values of 0.00, 2.50, and 1.25 mJ/m$^2$, respectively. **b**, Frequency spectra as a function of the $i$-DMI, when $J = 8.5 \times 10^7$ A/cm$^2$. Capital letters A, B, C, and D are linked to the phase diagram of Fig. 2.

**Figure 5 | Oscillation frequency vs. $i$-DMI.** Oscillation frequency as a function of $D$ for $J = 8.5 \times 10^7$ A/cm$^2$. The different states are indicated. The inset shows the energy related to ID transitions as a function of $D$ for $J = 8.5 \times 10^7$ A/cm$^2$.

**Figure 6 | Effect of the contact size on the NTD.** Frequency spectra as a function of the contact diameter $d_c$, when $J = 8.5 \times 10^7$ A/cm$^2$ and zero $i$-DMI, together with the spatial distribution of the topological density (red +1, blue -1) which corresponds to the frequency peak in the NTD dynamical state.

## Acknowledgements

This work was supported by the MIUR-PRIN 2010–11 Project 2010ECA8P3 "DyNanoMag" and the bilateral agreement Italy-Turkey project (Code B52I14002910005) "Nanoscale magnetic devices based on the coupling of Spintronics and Spinorbitronics". The authors thank Prof. Achim Rosch for the useful discussions.

## Author contributions

M. C. and R. T. performed micromagnetic simulations. R. Z. analyzed the data and wrote the paper. G. F, M. C., and R. T. conceived and designed the numerical experiment, analyzed the data and wrote the paper. All authors contributed to the general discussion and comment on the manuscript.

## Additional information

Supplementary information is available in the online version of the paper. Reprints and permissions information is available online at www.nature.com/reprints. Correspondence and requests for materials should be addressed to M.C. (mario.carpentieri@poliba.it).

## Competing financial interests

The authors declare no competing financial interests.

FIGURE 1

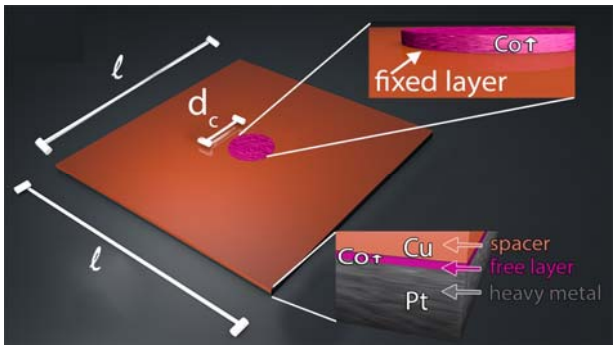

FIGURE 2

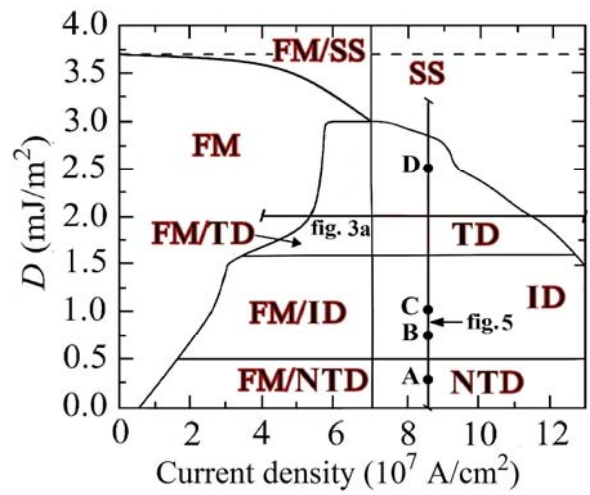

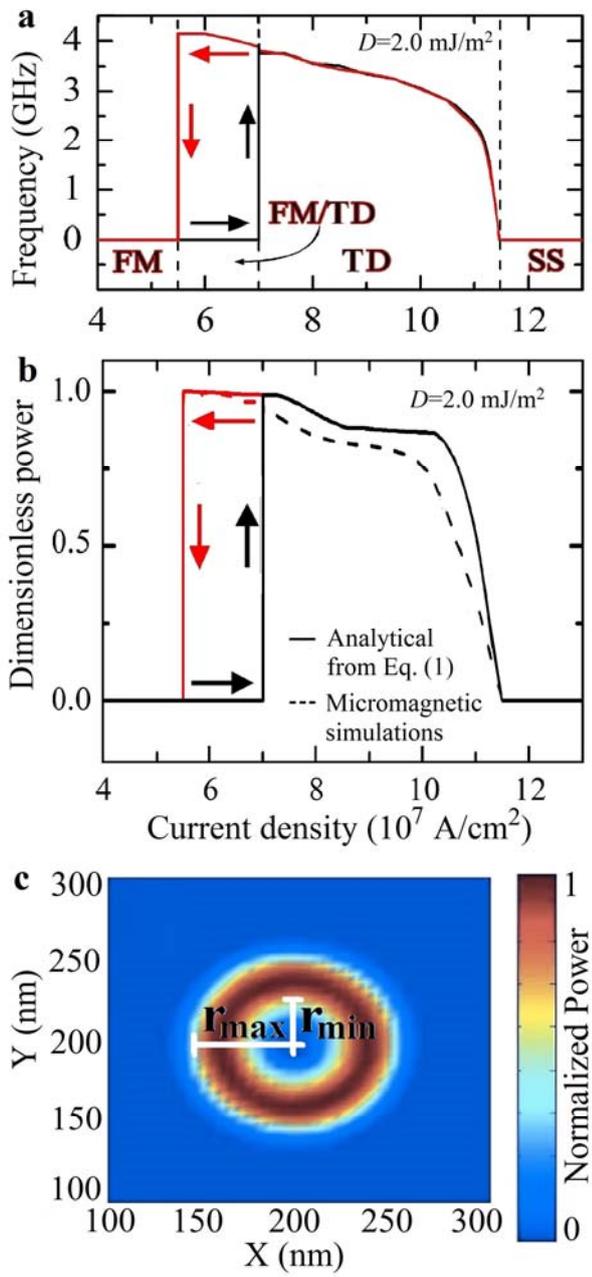

FIGURE 4

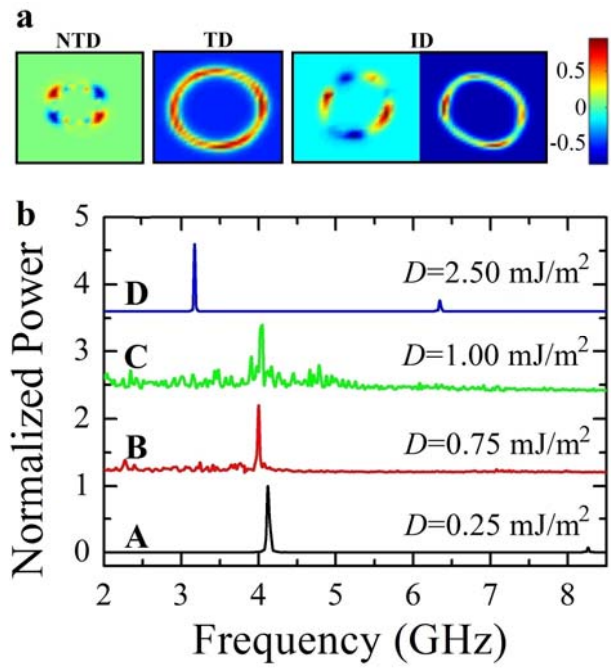

FIGURE 5

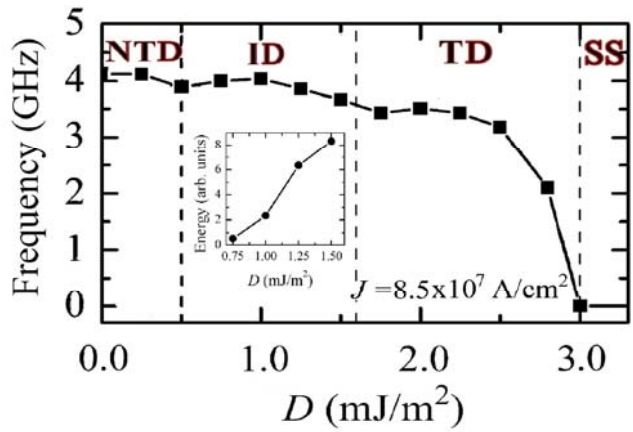